\documentclass[%
 reprint,
 amsmath,amssymb,
 aps,
]{revtex4-2}

\usepackage{graphicx}
\usepackage{dcolumn}
\usepackage{bm}
\usepackage{hyperref}

\begin{document}

\preprint{APS/123-QED}

\title{Colliding fields in two-dimensional dilaton gravity background}

\author{Mustafa Halilsoy}
 \altaffiliation[]{mustafa.halilsoy@emu.edu.tr}
\author{Chia-Li Hsieh}%
 \email{galise@gmail.com}
\affiliation{%
 Department of Physics, Faculty of Arts and Sciences, Eastern Mediterranean University, Famagusta,
North Cyprus via Mersin 10, Turkey 
}%


\date{\today}

\begin{abstract}
In the 1+1-dimensional dilatonic background of Callan et al. \cite{Callan1992}, we collide fields. Real and ghost fields are considered separately. The colliding ghost fields in the lineland creates particle-like wormholes. These structures, however, collapse to a black hole whenever the supporting ghosts are removed.
\end{abstract}

\maketitle



\section{\label{sec:level1}Introduction}

Dilatonic gravity in 1+1-dimensional (i.e. the “lineland”) provides a prototype model to understand and test a number of items that in standard 3+1-dimensions technical difficulties prevent us from \citep{Callan1992, Hayward1993, Hayward2002}.  It provides a simpler arena to test the non-linear field theory aspects such as quantum nature of vacuum, formation and collapse of blackholes/wormholes and their conversion from one type into the other. Formation of wormholes and their collapse into black holes has been analyzed in details \cite{Hayward2002}. Although the pure dilatonic vacuum in 1+1-dimensions is a completely solvable model, it provides yet a non-trivial model in field theory. In the model introduced by Callan et al. (CGHS) \cite{Callan1992}, the dilatonic vacuum of curved space is supported by a cosmological constant. Besides, the theory can have real and ghost fields to enrich the underlying dynamics.

In the present paper, we consider the same dilatonic background vacuum spacetime with a cosmological constant. Our purpose is to add incoming shock of real/ghost fields and obtain the resulting interacting region of fields. In other words, our background is a dilatonic vacuum rather than a flat vacuum of Minkowski spacetime. Owing to the mathematical simplicity of the model, the problem of colliding fields can also be solved without much effort. For details of the colliding waves in 3+1-dimensional general relativity, we refer to \cite{Griffiths2016}. Let us note that the problem of identifying plane waves and studying their collision in modified theories has not been considered extensively. Exceptional works are colliding waves in $F(R)=R^{N} (N>1)$ gravity in 3+1-dimensions \cite{Tahamtan2016},which shows much similarity to the colliding electromagnetic waves in Einstein$’$s gravity \cite{Bell1974}\cite{Halilsoy1988}. Due to the inherent non-linearity involved, which is less stringent in general relativity, finding exact solutions to such problems remain challenging. Newly we have also obtained exact colliding wave solutions of scalar and electromagnetic shock waves in 2+1-dimensions\cite{Halilsoy2022}.

We consider first collision of real fields in the dilatonic background. Expectedly, the mutual focusing of the fields create spacetime singularity that traps all future directed geodesics. For the case of colliding ghost fields, it is observed that formation of a wormhole ensues. From physical standpoint this is interesting, because a particle model was described first by Einstein and Rosen \cite{Einstein1935} as a wormhole geometry. We recall also that as a quantum process it was shown by Breit and Wheeler \cite{Breit1934} that collision of two $\gamma$-ray photons can give rise to a pair of electron-positron ($e^+-e^-$). A recent experiment \cite{Adam2021} made observations of $e^+-e^-$ pairs, but it remained unsettled whether this arose as a result of real or virtual photons. Our study is entirely classical here, yet finding imprints of particles, namely wormholes as a result of colliding ghost fields seems remarkable.

Our paper is organized as follows. In section II, we review briefly the model of CGHS and derive the field equations to be used. We consider colliding fields in the dilatonic background in section III. We complete the paper with our conclusion in section IV.




\section{The model of CGHS}
In the double null coordinate $(u, v)$ where our notation follows \cite{Hayward2002}, which differs slightly from that of CGHS \cite{Callan1992}, the line element is given by

\begin{equation}
  ds^2=-2e^{2\phi}dudv.
\end{equation}

Although the metric function and dilaton field can be taken different, for technical simplicity we shall follow the same trend of \cite{Hayward2002} to choose them from the outset equal. Thus $\phi(u, v)$ represents both the metric function as well as the dilaton field. Such a choice reduces the problem to a technically tractable one. The action of the model is

\begin{equation}
 I=\int\sqrt{-g}dudv[e^{-2\phi}(R+4(\bigtriangledown\phi)^2+4\lambda^2)-\frac{1}{2}(\bigtriangledown{f})^2+\frac{1}{2}(\bigtriangledown{g})^2],
\end{equation}in which all functions depend at most on $u$ and $v$. R is the Ricci scalar and $\lambda$ is the cosmological constant. From the signs of the terms, it can be recognized that $f(u, v)$ and $g(u, v)$ represent real and ghost fields, respectively. The scalar curvature is given by $R=4e^{-2\phi}\phi_{uv}$, where our notation is such that sub $u/v$ implies partial derivatives. By introducing the new variable for the dilaton ( and the metric function) $r=2e^{-2\phi}$ and upon variational principle, the field equations are as follow

\begin{eqnarray}
f_{uv}=0=g_{uv},
\\
 r_{uv}=-4\lambda^2,
\\
r_{uu}=g_{u}^2-f_{u}^2,
\\
r_{vv}=g_{v}^2-f_{v}^2.
\end{eqnarray}

To recall the vacuum dilaton case $(f=g=0)$, the foregoing set of equations integrate as
\begin{equation}
 r(u, v)=2m-4\lambda^2uv,
\end{equation}
where $m>0$, is a positive constant that plays the role of mass. From the analogy with a Schwarzschild black hole, the energy expression
\begin{equation}
 E=\frac{r}{2}(1-\frac{(\bigtriangledown r )^2}{4\lambda^2r^2}),
\end{equation}
gives that $E=m$, justifying the interpretation of $m$ as the mass. The scalar curvature gives
\begin{equation}
 R=\frac{4m\lambda^2}{m-2\lambda^2uv},
\end{equation}
which shows that the hyperbolic arc $uv=\frac{m}{2\lambda^2} $ is a singularity. The case $m=0$, is obviously a flat space that will not be considered. The detailed analysis that (6) makes a black hole can be found in \cite{Hayward1993}.


\section{Colliding waves in  the dilaton background}

In this section we assume that there is a background (vacuum) dilaton field filling the 1+1-dimensional spacetime as described in section (2). In such a spacetime, we consider the collision of various field as follows.



\subsection{Collision of two real fields}

We choose first the ghost field to vanish $g(u, v)=0$, and the real field $f(u, v)$ is given in the separable form
\begin{equation}
 f(u, v)=-1+\cos^2(au\theta(u))+\cos^2(bv\theta(v)),
\end{equation}
Here $(a,b)$ are constants and $(\theta(u), \theta(v))$ are Heaviside step functions to separate the different spacetime regions. Once we choose $f(u,v)$ , we solve the field equation (3) in the form
\begin{equation}
 r(u, v)=A(u)+B(v)-4\lambda^2uv,
\end{equation}
where $A(u)$ and $B(v)$ depend on the different null coordinates and satisfy
\begin{eqnarray}
A_{uu}=-f_u^2,
\\
 B_{vv}=-f_v^2.
\end{eqnarray}
Upon integration , by choosing the integration constant, we obtain
\begin{equation}
\begin{split}
r(u, v)=2m-4\lambda^2uv+
 \frac{1}{32}(1-\cos(4au\theta(u))\\
 +\frac{1}{32}(1-\cos(4bv\theta(v))
 -
 \frac{1}{4}(a^2u^2\theta(u)+b^2v^2\theta(v)).
\end{split}
\end{equation}

We note that in the term, $-4\lambda^2uv$ , which is from the background dilaton field, the null coordinates do not have step functions, whereas the other terms coming from the colliding fields do have. The problem now can be formulated as a collision in the following form $(Fig(1))$.

\subsubsection{Region $I$, $(u<0, v<0)$}
\begin{equation}
 f(u, v)=+1,
\end{equation}
\begin{center}
  $r(u, v)=2m-4\lambda^2uv$.
\end{center}

\subsubsection{Region $II$, $(u>0, v<0)$}
\begin{equation}
 f(u)=\cos^2(au),
\end{equation}
$r(u, v)=2m-4\lambda^2uv+\frac{1}{32}(1-\cos(4au))-
 \frac{1}{4}a^2u^2$.
\subsubsection{Region $III$, $(u<0, v>0)$}
\begin{equation}
 f(v)=\cos^2(bv),
\end{equation}
$r(u, v)=2m-4\lambda^2uv+\frac{1}{32}(1-\cos(4bv))-
 \frac{1}{4}b^2v^2$.
\subsubsection{Region $IV$, $(u>0, v>0)$}
\begin{equation}
 f(u, v)=-1+\cos^2(au)+\cos^2(bv),
\end{equation}

\begin{equation}
\begin{split}
r(u, v)=2m-4\lambda^2uv+\frac{1}{16}
  -\frac{1}{32}(\cos(4au)+\cos(4bv))\\
  -\frac{1}{4}(a^2u^2+b^2v^2).
\end{split}
\end{equation}

All the forgoing expressions are exact, satisfying the boundary conditions at $u=v=0$. To the leading orders, we can make the following expansions in $u$ and $v$, in case a perturbation solution is needed for $ u <<1 $ and $v << 1$,
\begin{equation}
 f(u, v)\approx1-a^2u^2-b^2v^2+\frac{1}{3}(a^4u^4+b^4v^4)+...
\end{equation}
\begin{equation}
 r(u, v)\approx2m-4\lambda^2uv-\frac{1}{3}(a^4u^4+b^4v^4)+...
\end{equation}
To the order $O(u^2), O(v^2)$, this is a black hole.

\subsection{Collision of two hyperbolic ghost fields}

For this case, we choose the real field $f=0$, and the ghost field as
\begin{equation}
 g(u, v)=-1+\cosh^2(au\theta(u))+\cosh^2(bv\theta(v)),
\end{equation}
in which $(a, b)$ are again arbitrary constants. In analogy with the case (A), we integrate $r(u, v)$ with the appropriate integration constants to obtain $r(u, v)$ in the interaction region ($IV$) $(u>0, v>0)$, given by
\begin{equation}
\begin{split}
r(u, v)=2m-4\lambda^2uv-\frac{1}{16}\\
  +\frac{1}{32}(\cosh(4au)+\cosh(4bv)) -\frac{1}{4}(a^2u^2+b^2v^2).
\end{split}
\end{equation}
Up to certain signs, and hyperbolic functions instead of trigonometric ones, colliding ghosts has similarities to the case of real fields. Also a similar figure to Fig.1 can be set where $f(u, v)$ is replaced by $g(u, v)$. In the interaction region (Region $IV$), expansion to the lowest orders with $ u<<1$ and $v << 1$ gives
\begin{equation}
 g(u, v)\approx1+a^2u^2+b^2v^2+\frac{1}{3}(a^4u^4+b^4v^4)+...
\end{equation}
\begin{equation}
 r(u, v)\approx2m-4\lambda^2uv+\frac{1}{3}(a^4u^4+b^4v^4)+...
\end{equation}
which are comparable with the expansions of the colliding real fields. It is observed that to the order $O(u^2), O(v^2)$, we have a black hole in the interaction region.

\begin{figure}
  \centering
  \includegraphics[width=0.3\textwidth]{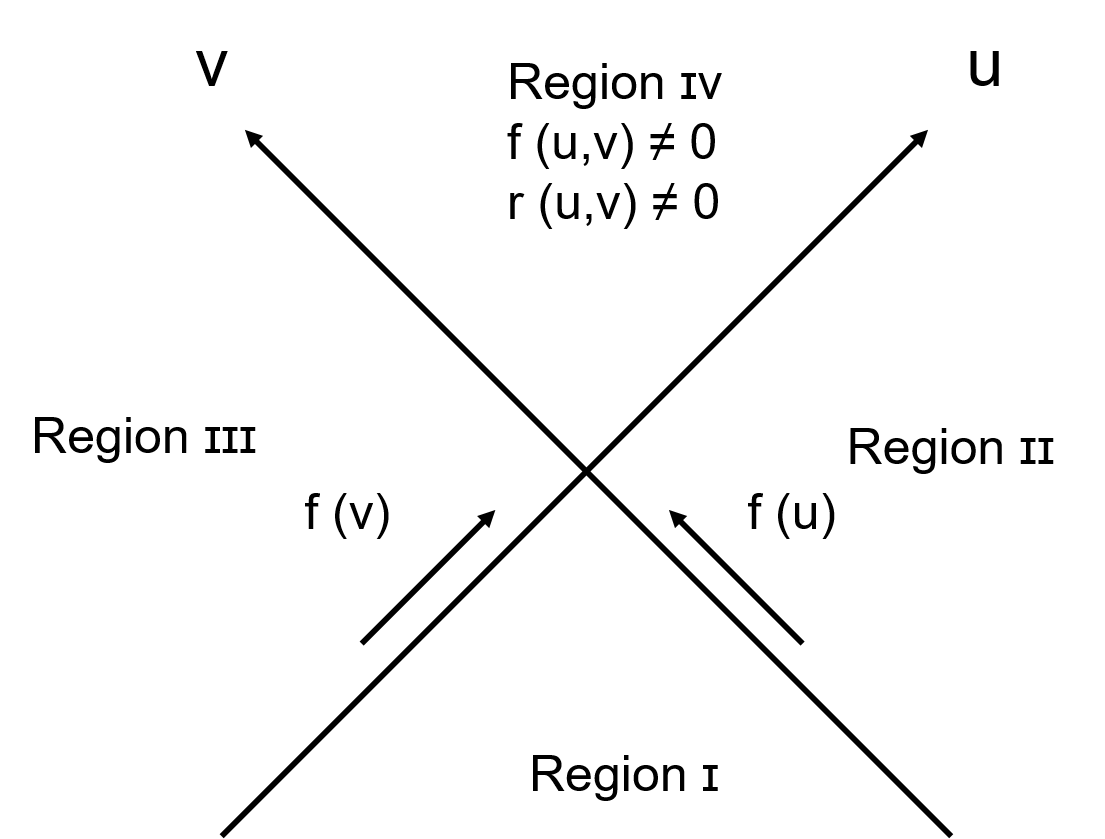}
  \caption{The picture describing the collision of two real fields. Region $I$ contains only a background dilaton, $r=2m-4\lambda^2uv$, with a trivial field $f=1$. Region $II$, has the incoming field $f(u)=\cos^2(au)$, which distorts also the background as given in the text. Region $III$, is the symmetric region of $II$, with $u\leftrightarrow v$ and $a\leftrightarrow b$. Region $IV$ is the interaction region having $f(u, v)$ and $r(u, v)$ with maximum terms that reduce to the appropriate incoming expressions. }\label{f1}
\end{figure}

\subsection{Colliding real and ghost fields}

With reference to Fig.1, we consider an incoming real field in Region $II$ given by
\begin{equation}
 f(u)=\cos^2(au\theta(u))
\end{equation}
and a ghost field in Region $III$ described by
\begin{equation}
 g(v)=\cosh^2(bv\theta(v)).
\end{equation}
These two fields interact at $u=v=0$ and develop into the Region $IV$, where the function $r(u, v)$ is given by
\begin{equation}
\begin{split}
 r(u, v)=2m-4\lambda^2uv+\frac{1}{32}(\cosh(4bv\theta(v))-\cos(4au\theta(u)))\\
 -\frac{1}{4}(a^2u^2\theta(u)+b^2v^2\theta(v)).
\end{split}
\end{equation}
It can be checked that this satisfies all boundary conditions and in particular the respective ghost and real field conditions
\begin{equation}
 r_{uu}=-f_u^2,
\end{equation}
\begin{equation}
 r_{vv}=g_v^2,
\end{equation}
are satisfied by the foregoing expressions Eg.s (26-28). An expression of $cosh()$ and $cos()$ functions reveal that to the second order we have
\begin{equation}
 r(u, v)\approx 2m-4\lambda^2uv+O(u^4, v^4),
\end{equation}
which is the black hole condition. In general, however, the expression (28) involves more complicated terms to argue in favor of a black hole.

\subsection{Colliding two linear ghost fields}

This type of ghost fields was already considered in \cite{Hayward2002}. We choose it here to discuss the problem of collision of such ghosts. We take
\begin{equation}
 g(u, v)=2\lambda(u\theta(u)-v\theta(v)),
\end{equation}
which implies that in the incoming regions we have linear fields with special coefficient. The field equations to be integrated are

\begin{eqnarray}
 r_{uv}=-4\lambda^2,
\\
r_{uu}=4\lambda^2\theta(u),
\\
r_{vv}=4\lambda^2\theta(v).
\end{eqnarray}
The integral for $r(u, v)$ is given in general by
\begin{equation}
 r(u, v)=2m-4\lambda^2uv+2\lambda^2(u^2\theta(u)+v^2\theta(v))+c_1u+c_2v,
\end{equation}
where $c_1$ and $c_2$ are arbitrary integration constants. We make choice for $c_1$ and $c_2$ and obtain the following two cases

\subsubsection{The choice of $c_1=c_2=0$, for $u>0, v>0$.}
This leads to
\begin{equation}
 r(u, v)=2m+2\lambda^2(u-v)^2,
\end{equation}
and by defining new coordinates
\begin{equation}
\sqrt{2}u=t+z,
\sqrt{2}v=t-z,
\end{equation}
we obtain
\begin{equation}
 r(z)=2m+4\lambda^2z^2,
\end{equation}
which is a wormhole solution with the throat $r=2m$ at $z=0$. The case $z>0$ and $z<0$ are the two mirror images of the wormhole. Thus, as a result of two colliding linear ghosts we obtain a wormhole.

\subsubsection{The choice of $c_1=c_2=-4\sqrt{m\lambda}$, for $u>0, v>0$.}
By inserting these constants into the integral (29), we obtain
\begin{equation}
 r(u, v)=2(\sqrt{m}-\lambda(u-v))^2,
\end{equation}
which gives the dilaton
\begin{equation}
 e^{-2\phi}=(\sqrt{m}-\lambda(u-v))^2,
\end{equation}
The line element reads now
\begin{equation}
 ds^2=\frac{-dt^2+dz^2}{(\sqrt{m}-\sqrt{2}\lambda z)^2},
\end{equation}
with the Ricci scalar $R=12\lambda^2$, which is regular. The two-dimensional dilation space can be expressed in the form
\begin{equation}
 ds^2=-e^{2\sqrt{2}\lambda Z}dt^2+dZ^2,
\end{equation}
where the new variable $Z(z)$ is defined by
\begin{equation}
Z(z)=-\frac{1}{\sqrt{2}\lambda}\ln(\sqrt{m}-\sqrt{2}\lambda z),
\end{equation}
which is again a wormhole solution in the new variable. It is seen that from both choices of the constants for $c_1$ and $c_2$, we obtain wormhole solutions as a result of colliding linear ghosts.

\subsection{Collapse of the wormhole}

In section (3.C), we describe how a linear dilation wormhole forms from the collision of two ghost fields. It can be easily checked that the fields
\begin{equation}
 g(u, v)=2\lambda(u\theta(-u)-v\theta(-v)),
\end{equation}
\begin{equation}
 r(u, v)=2m+2\lambda^2(u^2\theta(-u)+v^2\theta(-v)-2uv),
\end{equation}
also solves the field equations. We note that the present work differs from that of \cite{Hayward2002} by insertion of the step functions which makes the collision problem possible. As described in $Fig (2)$, the two ghosts collide first at $A$, forming a wormhole between $A$ and $O$. At $O$ the ghost fields vanish and this enforce the wormhole to collapse to a black hole. For completeness, let us add that between $A$ and $O$, a linear coordinate transformation on the null coordinate is required, which is out of our interest.


\begin{figure}
  \centering
  \includegraphics[width=0.3\textwidth]{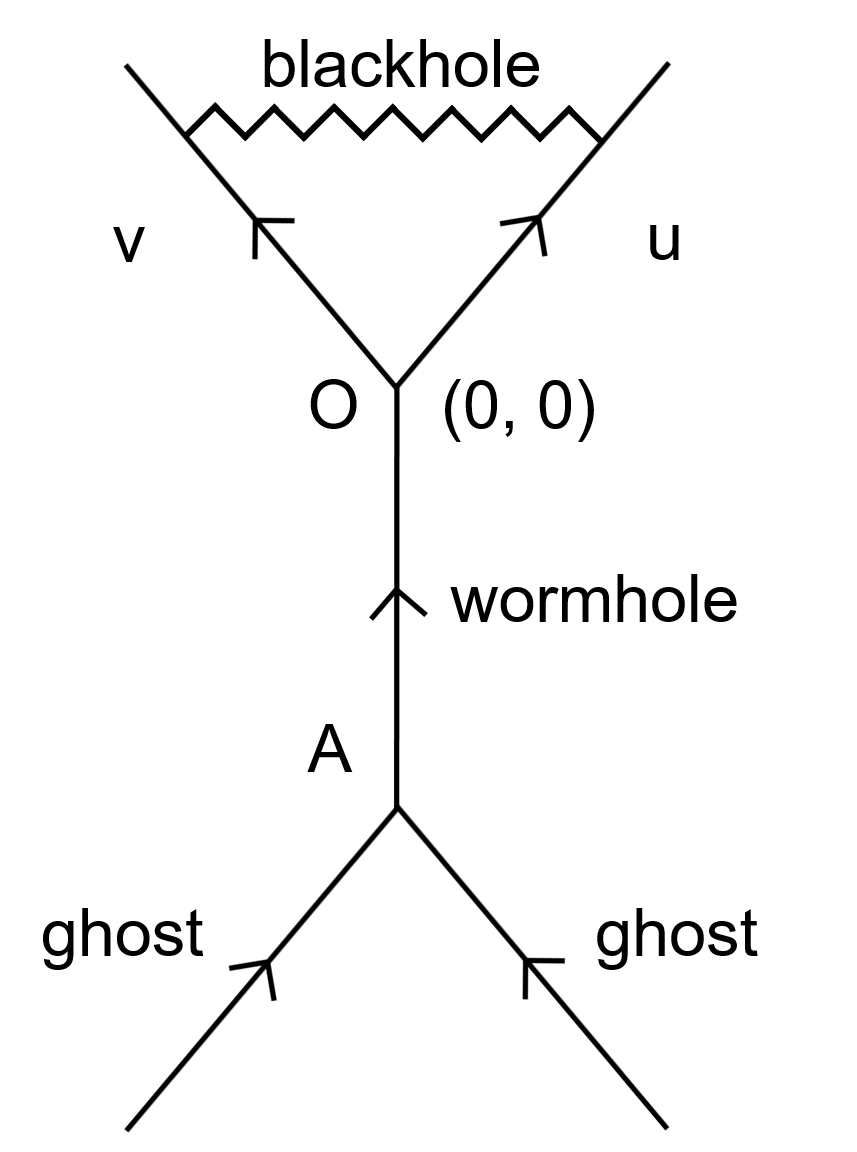}
  \caption{Diagram for creation and collapse of a wormhole. The line $AO$ represents a wormhole with throat $r=2m$, for, $u<0$, $v<0$ with $r=2m+2\lambda^2(u-v)^2$ as a yield of colliding ghosts in the past at $A$. At point $O$ the ghosts cease to support the wormhole and as a result the wormhole collapses to the black hole $r=2m-4\lambda^2uv$, once again.}\label{f2}
\end{figure}

\section{Conclusion}

As a toy model, the two-dimensional dilaton gravity model is revisited \cite{Callan1992}. In such a background that the only metric function and dilaton are identified, we considered the collision of real and ghost fields. As a matter of fact, when confined to one-dimension, it amounts to the collision of particles moving at the light speed. Even with this much simplification, the problem cannot be considered trivial. The dilaton constructed from the cosmological constant acts much like a catalyzer in these processes. It is shown that the ghost field creates a wormhole whereas real fields do not give interesting structure. It is not difficult also to add more fields and consider wave packets. We remind that the first wormhole idea was initiated as a geometrical model of an elementary particle \cite{Einstein1935}. This was due to the minimum radius which characterized a particle. Based on the dynamical equations we have shown that a wormhole radiates out two ghosts – which was created from – and collapses into a black hole. These processes all take place in 1+1-dimensional dilaton gravity model and to what extent such processes take place in higher dimensions remains to be seen. No doubt, it is natural to expect that collision of wormholes/blackholes also can be tackled easier in this low dimensional spacetime.

\end{document}